\documentclass[a4paper,11pt]{article}
\pdfoutput=1

\usepackage{a4}
\usepackage{jheppub}
\usepackage{graphics}
\usepackage{amssymb,amsmath}
\usepackage{color}
\usepackage{wasysym}
\usepackage[tight]{subfigure}
\usepackage{booktabs}

\newcommand{\beq}{\begin{equation}}
\newcommand{\eeq}{\end{equation}}
\newcommand{\beqn}{\begin{eqnarray}}
\newcommand{\eeqn}{\end{eqnarray}}

\newcommand{\sstop}{\tilde{t}_1}
\newcommand{\neut}[1]{\tilde{\chi}^0_{#1}}

\newcommand{\chapm}[1]{\tilde{\chi}^\pm_{#1}}
\newcommand{\mneu}[1]{m_{\tilde{\chi}^0_{#1}}}
\newcommand{\mcha}[1]{m_{\tilde{\chi}^\pm_{#1}}}
\newcommand{\mstop}[1]{m_{\tilde{t}_{#1}}}
\newcommand{\neu}[1]{\tilde{\chi}^0_{#1}}
\newcommand{\cha}[1]{\tilde{\chi}^\pm_{#1}}

\newcommand{\pb}{\ \mathrm{pb}}
\newcommand{\gev}{\ \mathrm{GeV}}
\newcommand{\tev}{\ \mathrm{TeV}}
\newcommand{\ifb}{\ \mathrm{fb}^{-1}}

\newcommand{\mathmet}{E_T^\mathrm{miss}}

\def\lsim  {\hspace{0.3em}\raisebox{0.4ex}{$<$}\hspace{-0.75em}\raisebox{-.7ex}{$\sim$}\hspace{0.3em}}

\preprint{DESY 13-052}
\title{Light stops emerging in $WW$ cross section measurements? }
\author[a]{Krzysztof Rolbiecki}
\author[b]{and Kazuki Sakurai}
\affiliation[a]{
Instituto de F\'{\i}sica Te\'{o}rica, IFT-UAM/CSIC,\\
C/ Nicol\'{a}s Cabrera, 13-15, Cantoblanco, 28049 Madrid, Spain}
\affiliation[b]{DESY,\\ Notkestrasse 85, D-22607 Hamburg, Germany}
\emailAdd{krzysztof.rolbiecki@desy.de}
\emailAdd{kazuki.sakurai@desy.de}
\abstract{
Recent ATLAS and CMS measurements show a slight excess in the $W^+W^-$ cross section measurement. While still consistent with the Standard Model within 1--2-$\sigma$, the excess could be also a first hint of physics beyond the Standard Model. We argue that this effect could be attributed to the production of scalar top quarks within supersymmetric models. The stops of $\mstop{1} \sim 200 \gev$ has the right pair-production cross section and under some assumptions can significantly contribute to the final state of two leptons and missing energy. We scan this region of parameter space to identify stop mass range preferred by the $W^+W^-$ cross section measurements. Taking one sample benchmark point we show that it can be consistent with low energy observables and Higgs sector measurements and propose a method to distinguish supersymmetric signal from the Standard Model contribution.
}

\begin{document}
\maketitle
\flushbottom


\section{Introduction\label{intro}}
The $W^+W^-$ diboson production process provides an important test of the electroweak (EW) interactions of the Standard Model (SM). Deviations from the SM predictions could arise due to new physics contributions, like anomalous triple gauge boson couplings or new particles decaying to the same final state as the electroweak gauge bosons.

The ATLAS and CMS experiments have performed measurements of the $W^+W^-$ pair production cross section in $pp$ collisions at $\sqrt{s} = 7\tev$ and $8\tev$ in the fully leptonic channel. Using the full dataset at 7~TeV, ATLAS measured the cross section $\sigma = 51.9\pm 2.0\ (\mathrm{stat}) \pm 3.9\ (\mathrm{syst}) \pm 2.0\ (\mathrm{lumi})\pb$~\cite{ATLAS:2012mec}, while quoting the SM prediction at next-to-leading (NLO) order of $\sigma = 44.7 \pm 2.0\pb$ at $\sqrt{s} = 7\tev$~\cite{Campbell:2009kg}. CMS measurements gave $\sigma = 52.4\pm 2.0\ (\mathrm{stat}) \pm 4.5\ (\mathrm{syst}) \pm 1.2\ (\mathrm{lumi})\pb$~\cite{CMS:2012cva}, compared to the SM expectation of $\sigma = 47.0 \pm 2.0\pb$~\cite{Campbell:2011bn}.\footnote{CMS and ATLAS use different methods to calculate the SM cross section, hence slightly different result.} At $\sqrt{s} = 8\tev$, only CMS has published the results using an integrated luminosity of $3.54\ifb$. It reported $\sigma =  69.9\pm 2.8\ (\mathrm{stat}) \pm 5.6\ (\mathrm{syst})
\pm 3.1\ (\mathrm{lumi})\pb$~\cite{Chatrchyan:2013oev} compared to the electroweak theory prediction of $\sigma = 57.3^{+2.4}_{-1.6} \pb$~\cite{Campbell:2011bn}. 

While the above results are far from being conclusive, there is a clear tendency at both experiments and center-of-mass energies for a slightly higher measured rate than the SM predictions. Interestingly, other EW measurements tend to be in a far better agreement with the SM than the $W^+W^-$ cross section measurement, see e.g.~\cite{Aad:2012twa,Aad:2012awa,Chatrchyan:2012sga,Chatrchyan:2012bd,ATLAS-CONF-2013-021,ATLAS-CONF-2013-020}. This provokes us to speculate that the origin of the discrepancy could be attributed to physics beyond the Standard Model (BSM). Based on lepton kinematic distributions, ATLAS~\cite{ATLAS:2012mec} imposes stringent limits on the anomalous $WWZ$ and $WW\gamma$ couplings. This leaves us with an exciting possibility of new particles being produced that contribute to the same final state --- two leptons and missing transverse energy --- as $W^+W^-$ pairs.

Production of supersymmetric (SUSY) particles could significantly affect measurement of $W^+W^-$ cross section in the fully leptonic final state. It was suggested in ref.~\cite{Curtin:2012nn} that in scenarios with charginos as the next-to-lightest supersymmetric particle one could expect an excess in the $W^+W^-$ cross section measurement, while avoiding constraints from searches in other channels. However, since the chargino pair-production cross section quickly decreases with the chargino mass, the size of enhancement is limited by the lower LEP limits~\cite{Beringer:1900zz} on the chargino mass. Nevertheless, the chargino contribution can be significant and would allow to decrease the tension between the prediction and the measurements, provided charginos are light and close to the existing bound, $\mcha{1} \gtrsim 100\gev$.

The other example of supersymmetric process that could contribute to the $W^+W^-$ cross section measurement is pair production of top squarks, as we argue in this paper. Light stops, motivated by naturalness argument~\cite{Kitano:2006gv,Asano:2010ut,Papucci:2011wy,Delgado:2012eu}, are extensively searched for at the LHC, see e.g.~\cite{ATLAS-CONF-2013-037,ATLAS-CONF-2013-048,ATLAS-CONF-2013-053,CMS-PAS-SUS-13-007,CMS-PAS-SUS-13-011} and references therein. Cross section is not a limiting factor here --- for $\mstop{1} \sim 200 \gev$ it easily exceeds $10\pb$. On the other hand, since stops decay hadronically one has to suppress the number of jets in the final state, in order to contribute to the leptonic final state without jets. This can be achieved by placing a chargino with a mass only slightly lower than the stop mass. The $b$-jets produced in the two-body stop decay, $\sstop \to \cha{1} b$, would be then too soft to be reconstructed. The chargino would further decay with on- or off-shell $W$, 
contributing to the dilepton final state,
\begin{equation}\label{eq:decay}
 \sstop \to \cha{1}\; b \to \neu{1}\; W^{(*)}\; b \to \neu{1}\; \ell \; \nu \; b\;,  
\end{equation}
where the $\neu{1}$ is the lightest supersymmetric particle (LSP) and escapes undetected.

The other possibility could be provided by three- or four-body stop decays where kinematics also limits $p_T$ of $b$-jets. These decay modes have been investigated in refs.~\cite{Yu:2012kj,Krizka:2012ah,Delgado:2012eu}, where the limits on the stop-neutralino parameter space have been derived using existing LHC analyses.
The stop production with a subsequent two-body decay is on the other hand constrained by a dedicated ATLAS study~\cite{ATLAS-CONF-2013-048}. However, because of the applied $m_{T2}$ cut, sensitivity of this search does not significantly affect a part of parameter space where $W$ becomes off-shell. Therefore, in section~\ref{sec:3} we fit the signal of the stop pair production, followed by the decay chain eq.~\eqref{eq:decay}, in order to find the minimal supersymmetric standard model (MSSM) parameters compatible with the $W^+W^-$ cross section measurement.

The paper is organised as follows. In the next section we briefly discuss the $W^+W^-$ cross section measurements, the relevant top squark search and simulation procedure. In section~\ref{sec:3} we perform a scan of the stop-neutralino masses to find a region consistent with the $W^+W^-$ excess and discuss a method to distinguish SUSY signal from SM processes. Finally, we conclude in section~\ref{conclusions}.  

\section{$WW$ and stop searches\label{sec:2}}
Both ATLAS and CMS have published $W^+W^-$ pair-production cross section measurements. ATLAS measured the $W^+W^-$ production cross section in $pp$ collisions at $\sqrt{s} = 7\tev$~\cite{ATLAS:2012mec}, while CMS published results for $\sqrt{s} = 7\tev$~\cite{CMS:2012cva} and $8\tev$~\cite{Chatrchyan:2013oev} using $\mathcal{L}_{\mathrm{int}} = 4.92\ifb$ and $3.54\ifb$, respectively. As discussed in Introduction, in both cases there was an excess in the observed number of events compared to the SM prediction. The experiments were looking at the leptonic channel, where the final state consists of two oppositely charged leptons (the same or opposite flavour) and missing transverse energy, $\ell^+ \ell^- + \mathmet$. In the following we briefly recapitulate the ATLAS and CMS searches.    

The main SM backgrounds for $pp \to W^+ W^- \to \ell^+ \ell^- \nu \bar{\nu}$ process originate from top quark production, Drell-Yan processes and other diboson pair production. In order to suppress top quark contribution a jet veto is applied. An event is rejected if there is at least one jet with $p_T > 25 (30) \gev$ in ATLAS (CMS) search. Drell-Yan production is suppressed using a cut on the invariant lepton mass, $m_{\ell\ell}$, and a \emph{projected (relative)} $E_{T,\mathrm{rel}}^{\mathrm{miss}}$ defined as
\begin{equation}
    E_{T,\mathrm{rel}}^{\mathrm{miss}}=\left\{ \begin{array}{ll}
                                           E_{T}^{\mathrm{miss}}\times \sin \Delta\phi_{\ell,j}& \qquad \mathrm{if}\ \Delta\phi_{\ell,j} < \pi/2 \\
                                           E_{T}^{\mathrm{miss}} & \qquad \mathrm{if}\ \Delta\phi_{\ell,j} \geq \pi/2
                                          \end{array}\;,\right.
\end{equation}
where $\Delta\phi_{\ell,j}$ is a difference in the azimuthal angle between $\mathbf{p}_T^{\mathrm{miss}}$ and the nearest lepton (jet).\footnote{ATLAS uses both jets and leptons to calculate this variable, while CMS only leptons.} After the cuts one obtains relatively clean sample of $W^+W^-$ events, with purity of $\sim 70\%$. The remaining background contribution is estimated using data-driven methods.\footnote{At this point the Higgs boson contribution, $h \to WW^*$, is not taken into account.}

Finally, we discuss the search for light stops performed by ATLAS~\cite{ATLAS-CONF-2013-048}, which covers a mass region relevant for our study. It targets the same final state as $W^+W^-$ analyses, two leptons with missing transverse momentum, but using a different set of cuts. Crucially, the signal regions in this study require $m_{\mathrm{T}2} > 90 \gev$. The $m_{\mathrm{T}2}$ variable~\cite{Lester:1999tx,Barr:2003rg} has a sharp kinematic edge at the $W$ boson mass for $t\bar{t}$ and $W^+W^-$ production. For the supersymmetric $\sstop \sstop^*$ production the kinematics could significantly differ from that of the top pair production, because of an additional contribution to missing transverse energy due to the LSPs. Therefore, stop production would populate a region of high $m_{\mathrm{T}2}$, where the SM backgrounds are suppressed. The situation changes for nearly-off-shell and off-shell $W$ in eq.~\eqref{eq:decay}. In this case, the $m_{\mathrm{T}2}$ cut will also result in suppression of the 
supersymmetric signal and loss of sensitivity. Since ATLAS presented search results for a similar scenario with $\mstop{1} - \mcha{1} = 10 \gev$ we can easily apply those exclusion bounds in our study. 

The same stop mass range is also constrained by another recent ATLAS search~\cite{ATLAS-CONF-2013-037}. The simplified models considered for interpretation of the results differ from those in ref.~\cite{ATLAS-CONF-2013-048} and the search targets one lepton plus jets final state. We find that the exclusion limit on our simplified model, discussed in the next section, is very similar to the one coming from~\cite{ATLAS-CONF-2013-048}.   

In order to find a range of stop parameters consistent with experimental searches we simulate events using \texttt{Herwig++~2.5.2}~\cite{Bahr:2008pv,Gigg:2007cr} with the default PDF set (MRST LO)~\cite{Martin:2002dr} and process them using fast detector simulation \texttt{Delphes~2.0.3}~\cite{Ovyn:2009tx}. We implement selection procedures and cuts for the relevant ATLAS and CMS searches discussed above. Furthermore, we validate the implementation by comparing efficiencies as reported by ATLAS and we find differences in efficiencies of less than 10\%.
Nevertheless, whenever possible we use the event rates of $W^+W^-$ and other SM processes given in the ATLAS and CMS publications. The stop signal is scaled to the NLO rate using \texttt{Prospino~2.1}~\cite{Beenakker:1997ut}. With this setup, we perform a scan described in the next section.   

\section{Stop contribution}\label{sec:3}

\subsection{Fitting a simplified model}

Given that the stop pair production events followed by the decay chain eq.~\eqref{eq:decay} 
contribute to the signal regions of the $W^+W^-$ measurements, the following questions should be addressed:
\begin{itemize}
\item  Which mass region can fit each experimental result well?
\item  Are those mass regions consistent with each other?
\item  Are those mass regions consistent with direct stop searches?
\item  How one can distinguish the stop contribution from genuine $W^+W^-$ events?
\end{itemize}
Postponing the last question to the next subsection, we address the first three in this subsection based on the simplified model approach.

Our simplified model considers exactly the same process as given by eq.~\eqref{eq:decay}.
As discussed in Introduction, the mass difference between the stop and chargino has to be small,
otherwise the $b$-quark from the stop decay would be reconstructed as a high-$p_T$ jet
and the event would be rejected by jet veto.
We therefore fix the chargino mass by $\mstop{1} - \mcha{1} = 10 \gev$.
With this assumption, the model is defined by two parameters: $\mstop{1}$ and $\mneu{1}$.
As mentioned in the previous section, ATLAS has recently presented the light stop search results 
using exactly the same simplified model.
Therefore, one can simply apply their exclusion limit to our simplified model parameter space. 

To find out which mass region fits the experimental results, we estimate the $\chi^2$ variable for each measurement as a function of the stop and neutralino masses:
\beq \label{eq:chi2}
\chi^2_i (\mstop{1}, \mneu{1}) = \frac{\left[ N^{(i)}_{\rm obs} - N^{(i)}_{\rm SM} - N^{(i)}_{\rm SUSY}(\mstop{1}, \mneu{1}) \right]^2 }{ \sigma_{i}^2 },
\eeq
where $ i $ specifies the measurement ($ i =\; $ATLAS7~\cite{ATLAS:2012mec}, CMS7~\cite{CMS:2012cva}, CMS8~\cite{Chatrchyan:2013oev}), 
$N^{(i)}_{\rm obs}$ is the number of observed events in the signal region,
$N^{(i)}_{\rm SM}$ and $N^{(i)}_{\rm SUSY}$ are the predicted contributions
from the Standard Model and SUSY, respectively.
The total uncertainty, $\sigma_i$, includes the systematic and statistical uncertainties taken from \cite{ATLAS:2012mec, CMS:2012cva, Chatrchyan:2013oev} as well as the uncertainty of 15\% on the stop cross section, see ref.~\cite{Kramer:2012bx}.
We add those uncertainties in quadrature: $\sigma_i^2 = \sigma_{\rm syst}^2 + \sigma_{\rm stat}^2 + \sigma_{\tilde t}^2 $, where $\sigma_{\tilde t} = 0.15 \cdot N_{\rm SUSY}^{(i)}$.  
The $N^{(i)}_{\rm SM}$ includes not only the $W^+W^-$ contribution but also the other SM contributions such as $t \bar t$ and $h \to WW^*$ processes.\footnote{The SUSY-EW contribution from a direct chargino and neutralino production followed by leptonic decays is model dependent and, in any case, factor 20--50 smaller than the stop pair production and, therefore, can be neglected.} All the factors, except for the $N^{(i)}_{\rm SUSY}$, are provided in refs.~\cite{ATLAS:2012mec, CMS:2012cva, Chatrchyan:2013oev}. 

We estimate $N^{(i)}_{\rm SUSY}(\mstop{1}, \mneu{1})$ in the following procedure.
We generate a grid in the ($\mstop{1}, \mneu{1}$) plane with a $10\gev \times 10\gev$ step size.     
In each grid point, $10^5$ events of $\tilde t_1 \tilde t^*_1$ followed by the decay eq.~\eqref{eq:decay} 
are generated.
We then apply the cuts used in the $W^+W^-$ cross section measurement 
and estimate the efficiency, $\epsilon_i(\mstop{1}, \mneu{1})$.
The NLO cross section of the stop pair production, $\sigma_{\tilde t}(\mstop{1})$, 
is calculated using \texttt{Prospino~2.1}~\cite{Beenakker:1997ut}. 
Finally, the SUSY contribution to the signal region is obtained 
by $N^{(i)}_{\rm SUSY}(\mstop{1}, \mneu{1}) = {\cal L}_{\rm int} \cdot \sigma_{\tilde t}(\mstop{1}) 
\cdot [{\rm BR}(\tilde t_1 \to \ell \nu \neut{1})]^2 
\cdot \epsilon_i(\mstop{1}, \mneu{1})$, where ${\cal L}_{\rm int}$ is the integrated luminosity.  
   
\begin{figure}[t]
\begin{center}
  \subfigure[]{\includegraphics[width=0.48\textwidth]{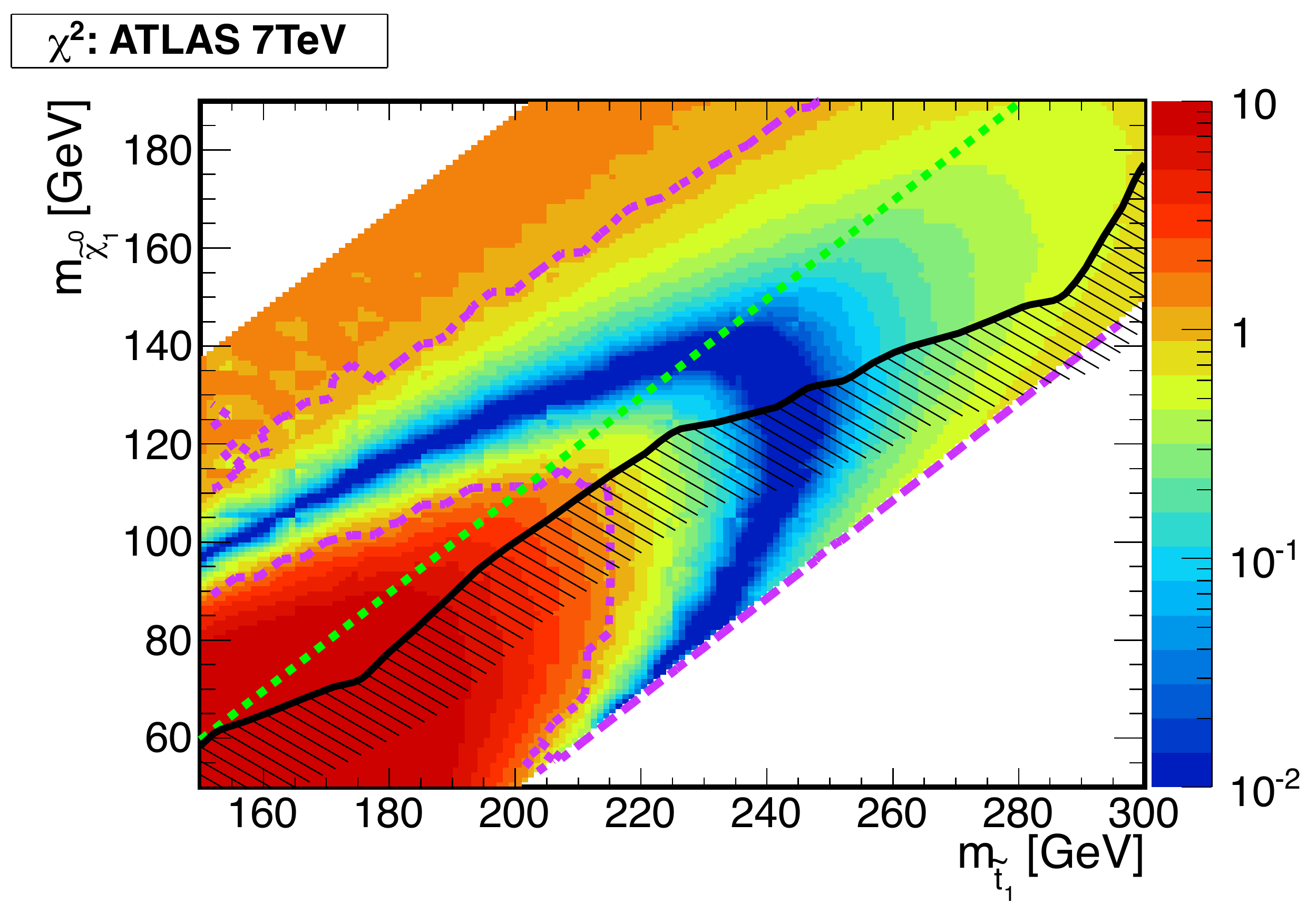}\label{fig:atlas7}} \hspace{0.2cm}
  \subfigure[]{\includegraphics[width=0.48\textwidth]{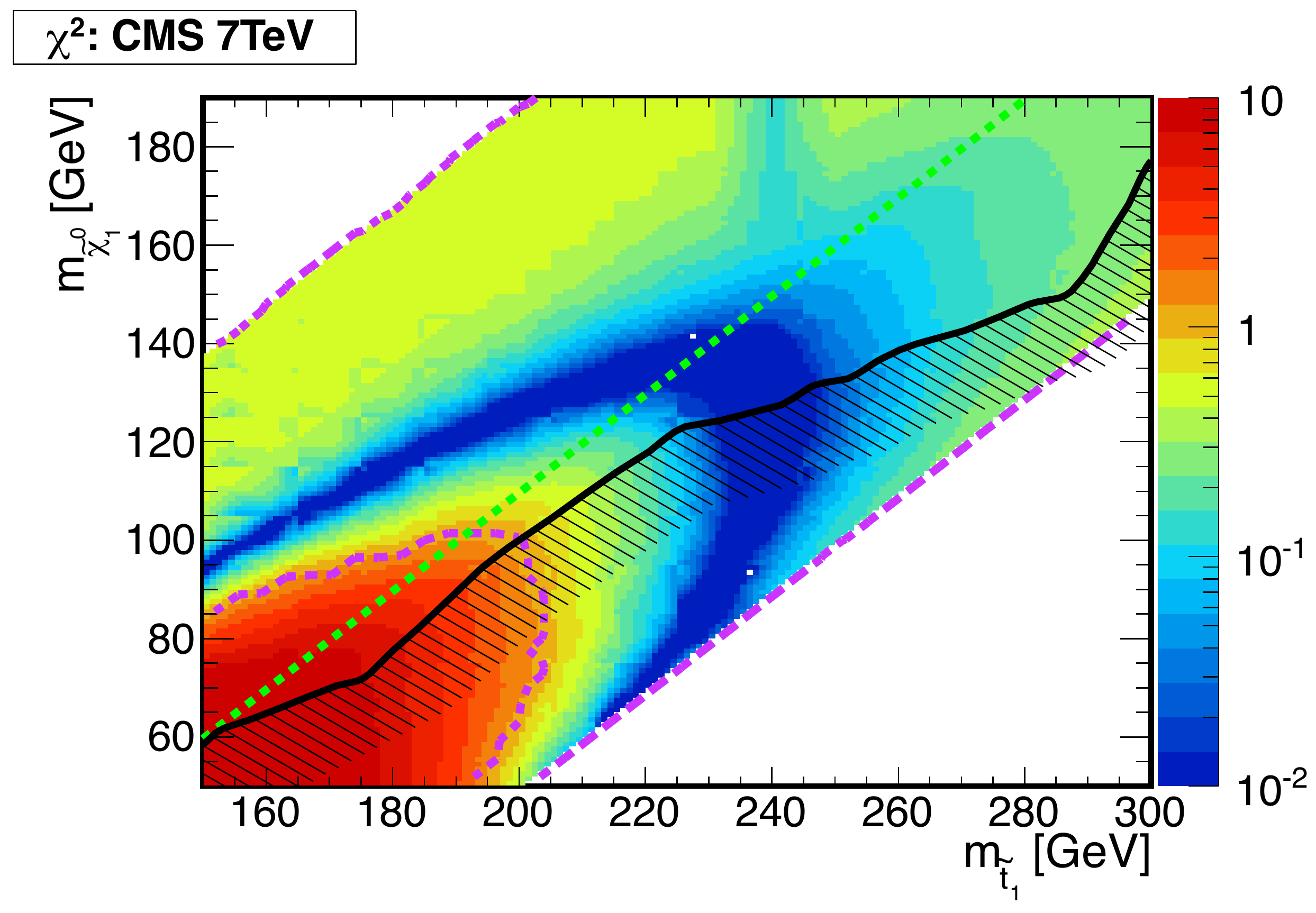}\label{fig:cms7}} 
  \subfigure[]{\includegraphics[width=0.48\textwidth]{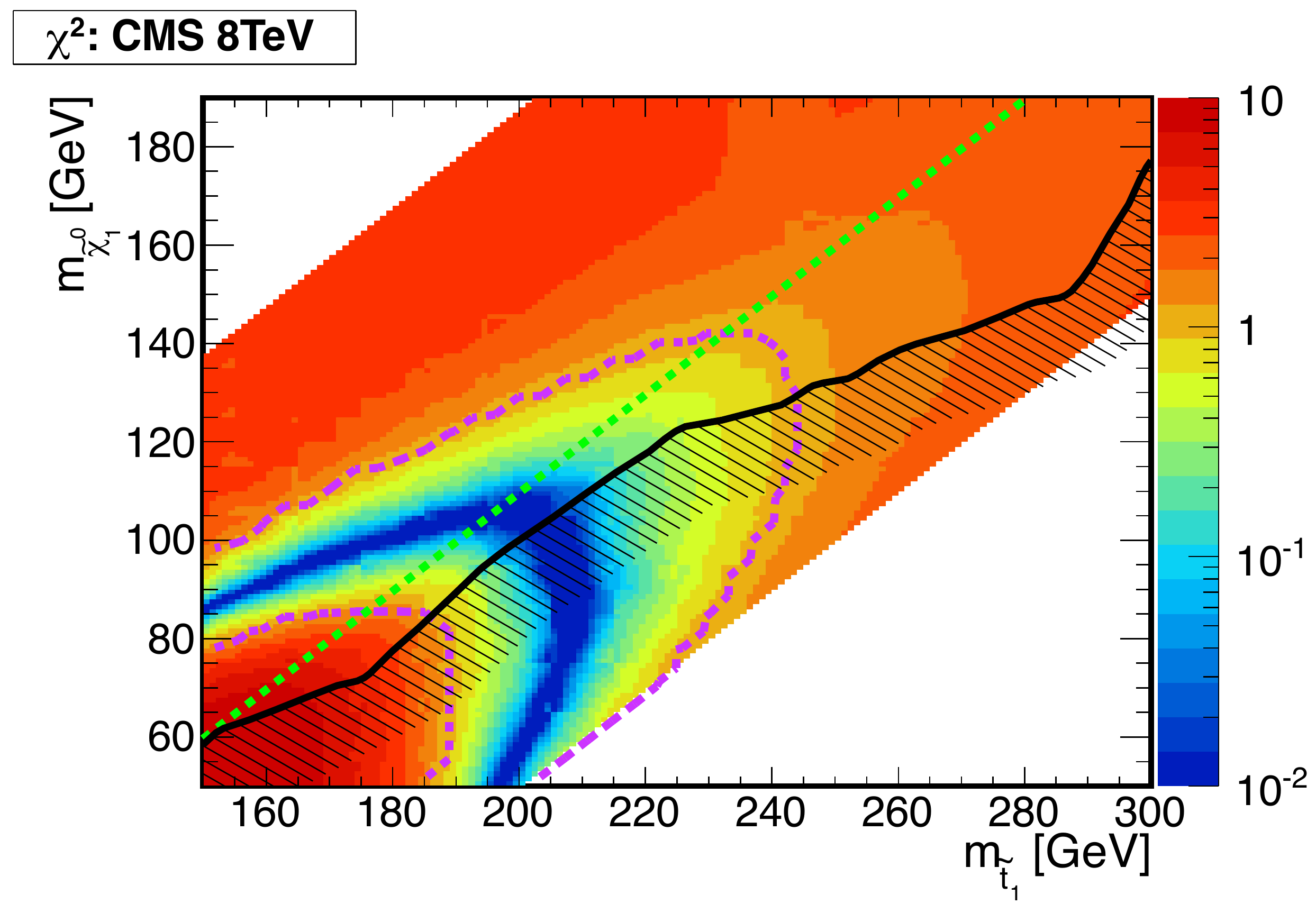}\label{fig:cms8}} \hspace{0.2cm}
  \subfigure[]{\includegraphics[width=0.48\textwidth]{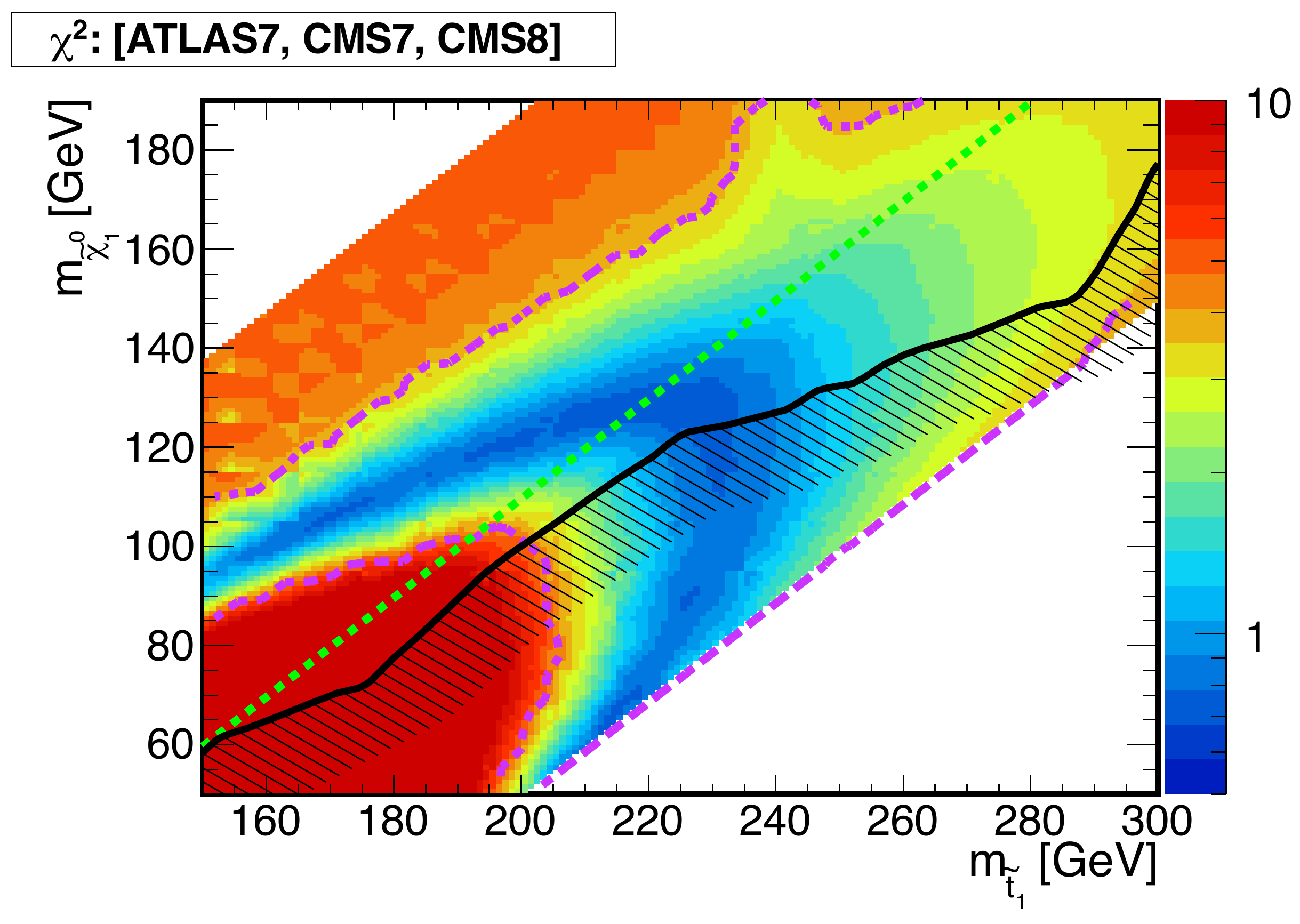}\label{fig:comb}} 
\caption{ The $\chi^2$, eq.~\eqref{eq:chi2}, distributions in the $(\mstop{1}, \mneu{1})$ plane for each of the measurements, ATLAS7, CMS7 and CMS8. In panel (d), the sum of  $\chi^2$s for the three measurements is shown. Blue areas represent the lowest values of $\chi^2$ and the region preferred by the experiments. A green dashed line indicates the kinematical threshold for $\chapm{1} \to W^\pm \neut{1}$ decay. The shaded region below a black line is excluded by the ATLAS direct search~\cite{ATLAS-CONF-2013-048}. A dashed purple line shows a $68\%$ CL region.    \label{fig:chisq} }
\end{center}
\end{figure}

Figures~\ref{fig:chisq}\,(a)--(c) show the $\chi^2$ in the ($\mstop{1}, \mneu{1}$) plane 
for the ATLAS7, CMS7 and CMS8 measurements, respectively. The area below a black line is excluded by the ATLAS direct stop search~\cite{ATLAS-CONF-2013-048}.
In the white top-left region chargino becomes the LSP. Near the boundary of the chargino LSP region, the leptons from the $\chapm{1} \to \ell \nu \neut{1}$ decay
become too soft to be detected, leading to $N^{(i)}_{\rm SUSY} \to 0$.  Therefore in the vicinity of the boundary
the $\chi^2$ approaches to the SM value.

As can be seen, the best fit regions of the three measurements form a similar arc-shaped area, 
which is roughly symmetric with respect to the dashed green line.
The dashed green line shows the kinematical threshold of the $\chapm{1} \to W^\pm \neut{1}$ decay.
In the region above this line, the $W$ becomes off-shell
and the lepton from the three-body decay, $\chapm{1} \to \ell \nu \neut{1}$, becomes softer as moving away from the line,
which in turn requires a smaller stop mass to compensate degradation of the efficiency by an enhancement of the cross section.
In the region below this line, the $W$ from the two-body decay, $\chapm{1} \to W \neut{1}$, becomes more 
energetic as moving away from the threshold.
This results in degradation of the efficiency, because the lepton and neutrino from the boosted $W$ decay 
are collimated, leading to a smaller projected $E_T^{\rm miss}$.
The neutralinos do not contribute much to the $E_T^{\rm miss}$, because in the near-threshold region 
they tend to be back-to-back in the transverse plane and their contributions cancel out.
In the opposite limit, $\mneu{1} \ll m_W$, most of the chargino momentum is carried by the $W$
and the neutralino becomes soft.

The dashed purple curves show the $68\%$~CL regions.
The regions are somewhat broad for ATLAS7 and CMS7.
In fact, the SM prediction agrees with the data within 1-$\sigma$ accuracy for CMS7,
therefore adding the stop contribution does not provide a meaningful improvement.
On the other hand, the 1-$\sigma$ region for CMS8 is much more localised around 
$\mstop{1} \lsim 250\gev$ and $80\gev \lsim \mneu{1} \lsim 140\gev$. 
This is because the discrepancy between the data and the SM is at about 2-$\sigma$ level
and a large stop contribution is required to account for the observed excess.
Interestingly, for each measurement a large part of the 1-$\sigma$ region is 
not excluded by the ATLAS light stop search~\cite{ATLAS-CONF-2013-048}. 
Moreover, the preferred regions from the three independent measurements are consistent with each other,
although two of those provide somewhat broad 1-$\sigma$ regions.
This agreement is nontrivial since the cuts and the center-of-mass energies are different in these measurements.
Figure~\ref{fig:comb} shows the sum of $\chi^2$ values for the three measurements.
As can be seen, a significant part of the preferred parameter region is consistent with the ATLAS light stop constraint.

We would also like to comment on the bottom left corner of the parameter space where the models are strongly disfavoured by the data.  In this region, the contribution from the $\tilde t_1 \tilde t^*_1$ events is too large.
This indicates that an analysis similar to the $W^+W^-$ cross section measurement can also be applied to the light stop search. 
In fact, the disfavoured region spreads to the yet unconstrained area.
A dedicated analysis along these lines would be able to extend the stop exclusion limits.

\begin{figure}[t]
\begin{center}
  \subfigure[]{\includegraphics[width=0.48\textwidth]{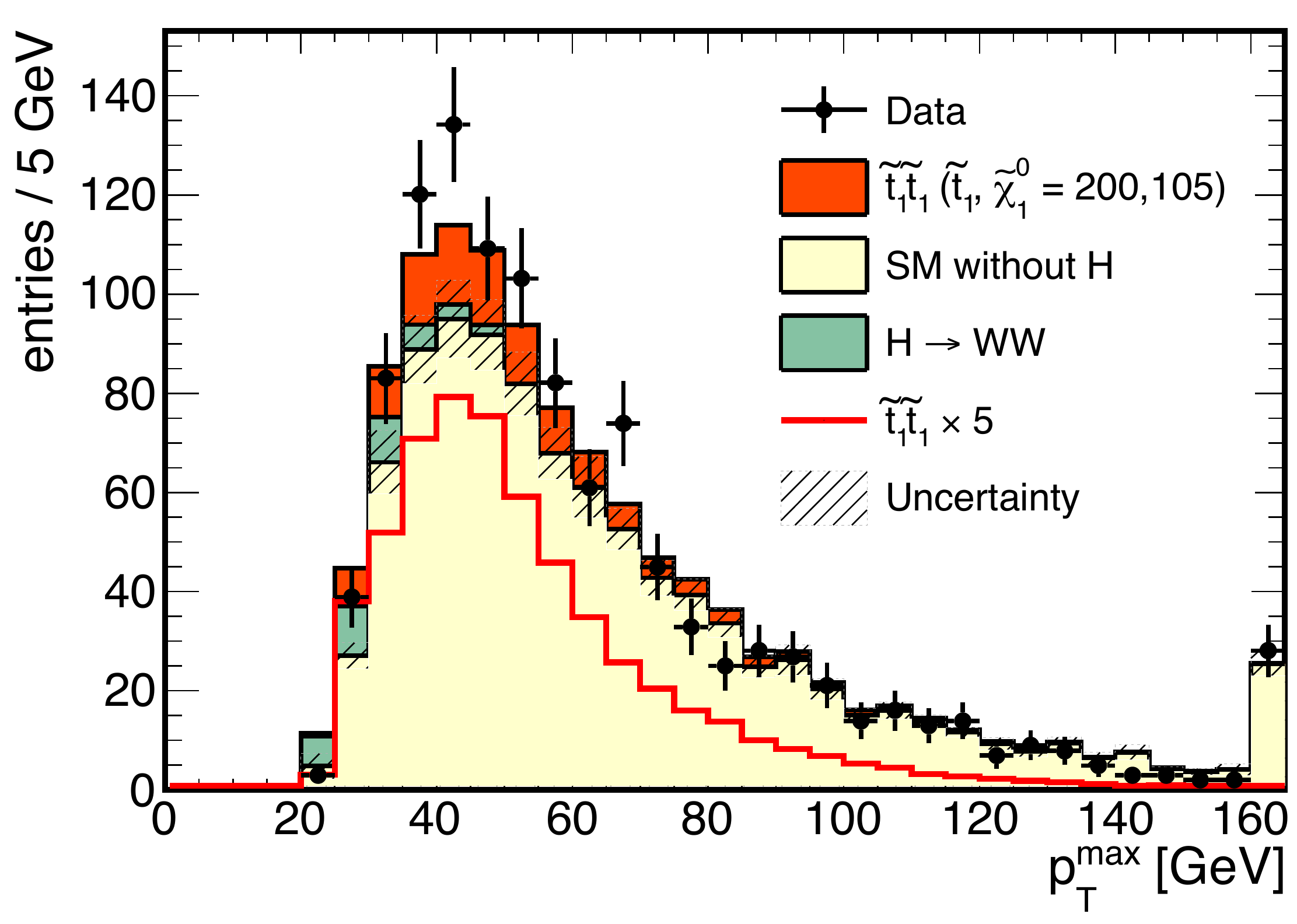}\label{fig:pt1}} \hspace{0.2cm}
  \subfigure[]{\includegraphics[width=0.48\textwidth]{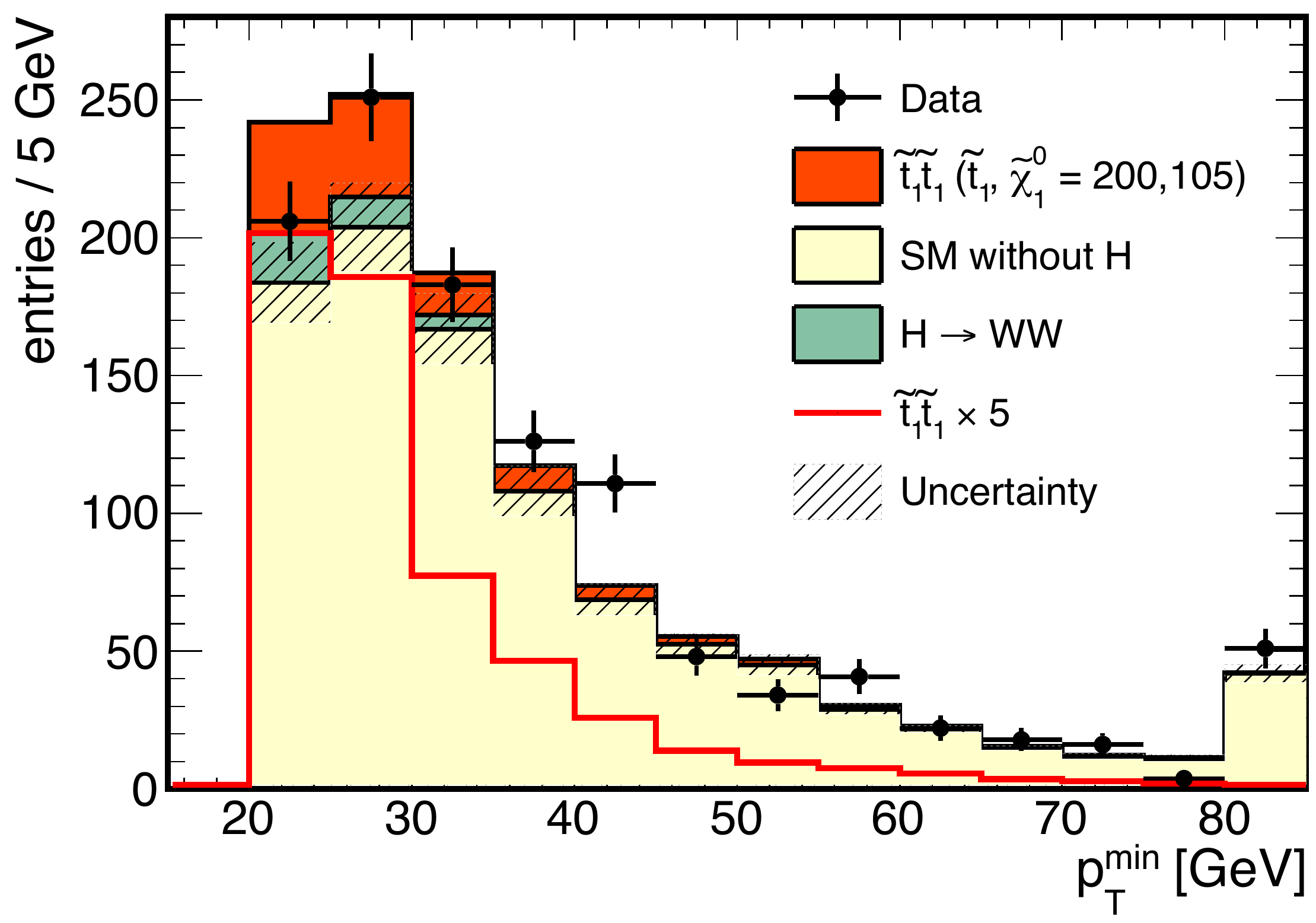}\label{fig:pt2}} 
  \subfigure[]{\includegraphics[width=0.48\textwidth]{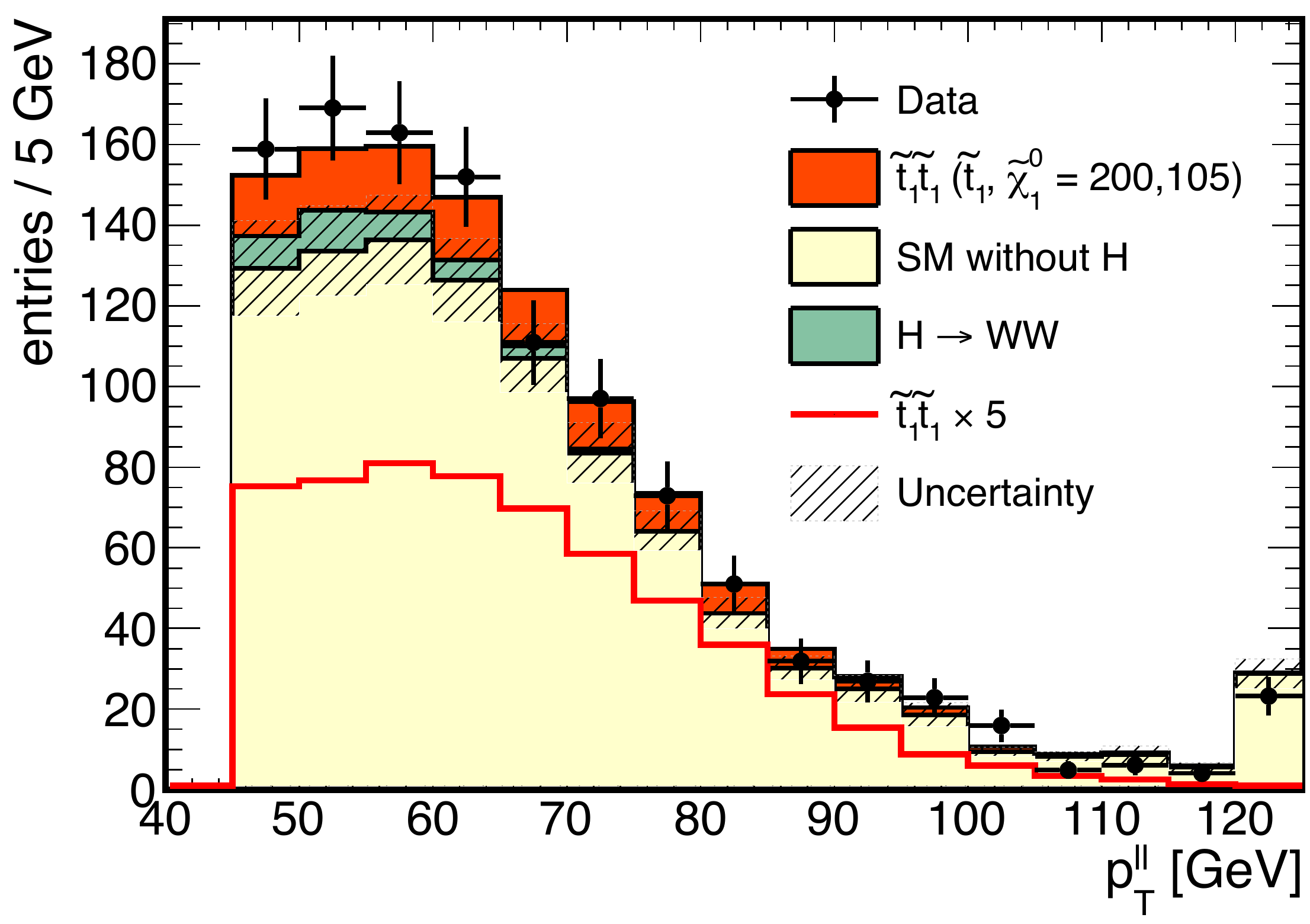}\label{fig:pt_ll}} \hspace{0.2cm}
  \subfigure[]{\includegraphics[width=0.48\textwidth]{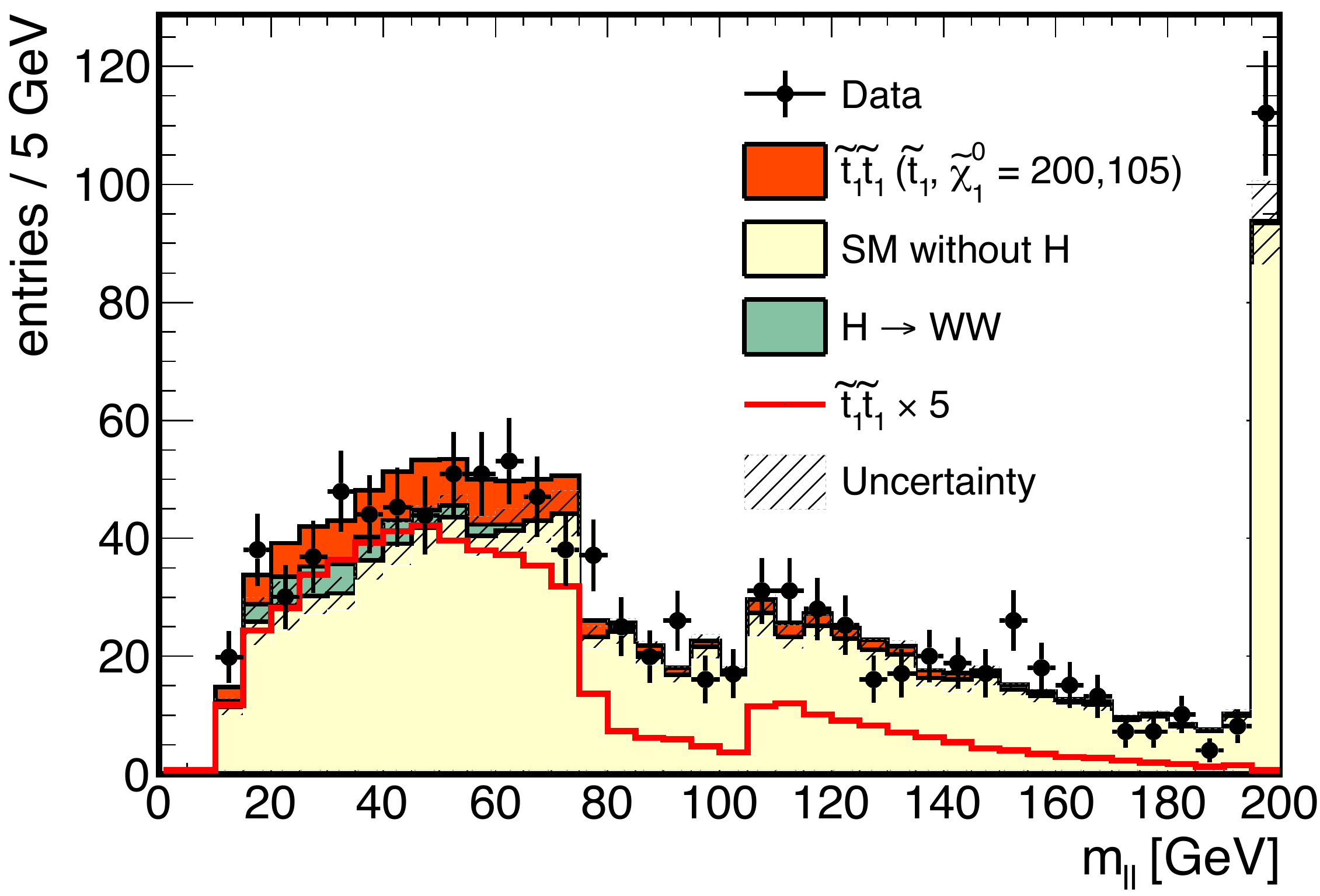}\label{fig:m_ll}} 
\caption{Distributions of: (a) the leading lepton transverse momentum $p_T^{\mathrm{max}}$, (b) the trailing lepton transverse momentum $p_T^{\mathrm{min}}$, (c) the dilepton system transverse momentum $p_T^{\ell\ell}$, and (d) the dilepton invariant mass $m_{\ell\ell}$. The SM, Higgs and stop contributions are shown separately. The genuine stop contribution is also depicted for comparison and multiplied by factor 5 for convenience. The SM event numbers, data points and uncertainties are taken from ref.~\cite{Chatrchyan:2013oev}. Note that we display the expected number of SM events, i.e.\ rescaled compared to figure~1 of ref.~\cite{Chatrchyan:2013oev}. We follow a convention proposed in ref.~\cite{Curtin:2012nn} in presenting this plot.  \label{fig:dist} }
\end{center}
\end{figure}

If we are indeed observing the stop contribution, the stop events can fit not only
the number of observed events after the cuts but also any observed distribution.
Therefore we compare the data and our light stop model 
to the distributions provided in ref.~\cite{Chatrchyan:2013oev}. 
Figure~\ref{fig:dist} shows the distributions of (a) the $p_T$ of the leading lepton, (b) the $p_T$ of the trailing lepton,
(c) the $p_T$ of the dilepton system and (d) the dilepton invariant mass.
We choose $(\mstop{1}, \mneu{1}) = (200, 105)\gev$ as a benchmark point. The NLO stop cross sections are $11.3\pb$ and $17.3\pb$ at $\sqrt{s} = 7$ and 8~TeV, respectively. The stop signal efficiency for the CMS selections~\cite{Chatrchyan:2013oev} is $\sim 0.18\%$ and we expect a contribution of about 110 events at $\sqrt{s} = 8\tev$ with the integrated luminosity $3.54\ifb$. The histograms show a good agreement between the data and the light stop model.
The shapes of the SM and the light stop contributions are very similar,
therefore distinguishing between them would be very difficult 
when using only the provided kinematic distributions. The histograms of the remaining two measurements also show a good level of agreement after including the stop signal.
In the next subsection, we propose a method to distinguish the stop contribution 
from the SM.

The best fit point can be easily realised within the MSSM. With one of the stop states heavy, one needs a large splitting between the left and right stops to obtain the Higgs boson mass in agreement with experiment~\cite{Aad:2012tfa,Chatrchyan:2012ufa}. We fix the stop sector by choosing: $\mstop{R} = 195 \gev$, $\mstop{L} = 2000 \gev$ and $A_t = 2000 \gev$. The chargino and neutralino sectors are given by: $M_1 = 105 \gev$, $M_2 = 190 \gev$, $\mu = 2500 \gev$ and $\tan\beta = 15$. Masses of other sfermions,  Higgs bosons and gluino are fixed by: $M_{\rm SUSY} = M_3 = M_{A^0} = 2000 \gev$, except for the mass of the right bottom squark, $m_{\tilde{b}_R} = 1000 \gev$. We do not include off-diagonal entries in the sfermion mass matrices in the super-CKM basis. For such a choice of parameters we obtain:
$\mstop{1} = 203.7 \gev$, $\mneu{1} = 104.9 \gev$ and $\mcha{1} = 189.5 \gev$, in the region preferred by the fit. Using \texttt{FeynHiggs~2.9.4}~\cite{Heinemeyer:1998np,Heinemeyer:1998yj,Degrassi:2002fi,Frank:2006yh} we evaluated the Higgs boson mass to be $m_h = 125.6 \gev$, while the rate in $h\to \gamma \gamma$ mode turns out to be $R_{\gamma\gamma} = 1.05\cdot R_{\gamma\gamma}^{\rm SM}$ compared to the SM value. Low energy observables have been checked with \texttt{SuperIso~3.3}~\cite{Mahmoudi:2007vz,Mahmoudi:2008tp}: $\mathrm{BR}(B\to X_s \gamma) = 3.7 \times 10^{-4}$ and $\mathrm{BR}(B_s \to \mu\mu) = 3.45 \times 10^{-9}$, and are consistent with the current experimental values~\cite{Amhis:2012bh,Aaij:2012nna}.   

\subsection{Stop's smoking gun}
If the excess in the $W^+W^-$ cross section measurement is confirmed with a higher significance, it will be crucial to confirm that it indeed originates from beyond SM physics. Therefore, we discuss here an angular distribution that could help to discriminate between the SM contribution and supersymmetric origin. As a working point we choose the benchmark scenario discussed in the previous subsection: $m_{\tilde{t}_1} = 200 \gev$, $\mcha{1} = 190 \gev$ and $\mneu{1} = 105 \gev$.

Due to different spins and production mechanisms of $W$ bosons and top squarks one can expect differences in the polar angle distribution, $\cos\theta^*$, of initially produced particles in the hard process center-of-mass frame, as discussed in refs.~\cite{Barr:2005dz,Choi:2006mr,Alves:2007xt,MoortgatPick:2011ix}. This indeed is the case as can be seen in figure~\ref{fig:costh_raw}, where $W^+W^-$ production exhibits a strong enhancement in the forward direction. In case of stops, the effect is much less pronounced even though the forward direction is also preferred. A similar behaviour can be observed for $t\bar{t}$ production also shown in the figure. As discussed in ref.~\cite{MoortgatPick:2011ix}, such a difference could affect angular distributions of the final state particles and provide a strong discrimination between different models.

In order to probe the production distribution more directly, we use the following observable~\cite{Barr:2005dz}:
\begin{equation}\label{eq:costhstar}
\cos\theta_{\ell\ell}^* = \tanh\left(\frac{\Delta \eta_{\ell\ell}}{2}\right) \;, \qquad \Delta\eta_{\ell\ell}= \eta_{\ell_1} - \eta_{\ell_2}\;,
\end{equation}
where $\Delta  \eta_{\ell\ell}$ is the difference of the pseudorapidities between the leading and the trailing lepton. This variable is the cosine of the polar angle of the leptons with respect to the beam axis in the frame where the pseudorapidities of the leptons are equal and opposite. Being a function of the difference of pseudorapidities, it is longitudinally boost-invariant. Figure~\ref{fig:cth0} shows $\cos\theta_{\ell\ell}^*$ distribution for $W^+W^-$, $t \bar t$ and $\tilde{t}_1 \tilde{t}_1^*$ pairs. Much of the difference seen in figure~\ref{fig:costh_raw} is now absent, which makes distinction between the two processes significantly more difficult. As pointed out in ref.~\cite{Barr:2005dz}, the $\cos\theta_{\ell\ell}^*$ observable requires high boosts of initially produced particles. However, both $W^+W^-$ and stops have a significant fraction of events produced close to the threshold, that partially dilutes the expected difference in the final state distribution. 

\begin{figure}[t]
\begin{center}
  \subfigure[]{\includegraphics[clip,width=0.48\textwidth]{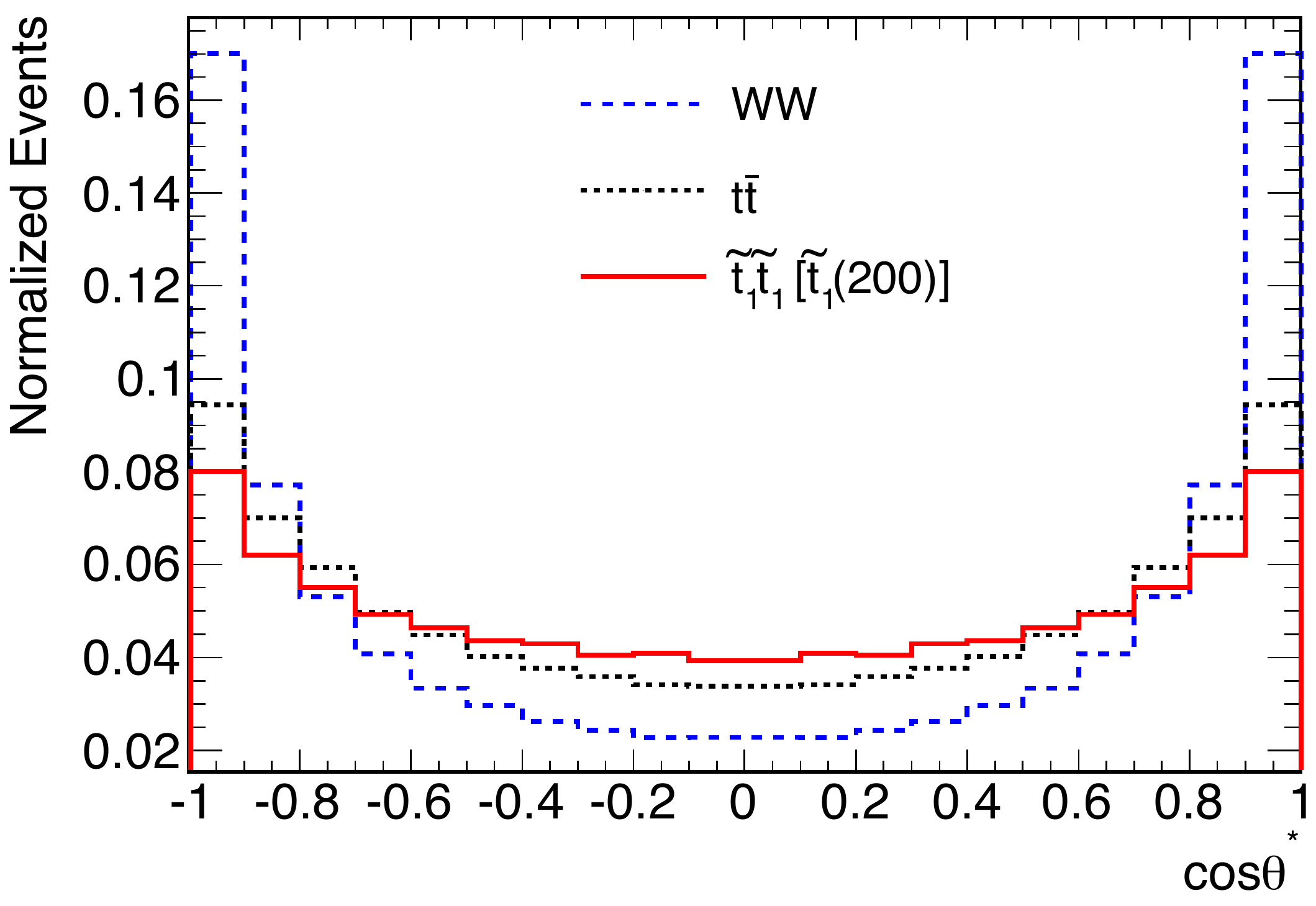}\label{fig:costh_raw}} \hspace{0.2cm}
  \subfigure[]{\includegraphics[clip,width=0.48\textwidth]{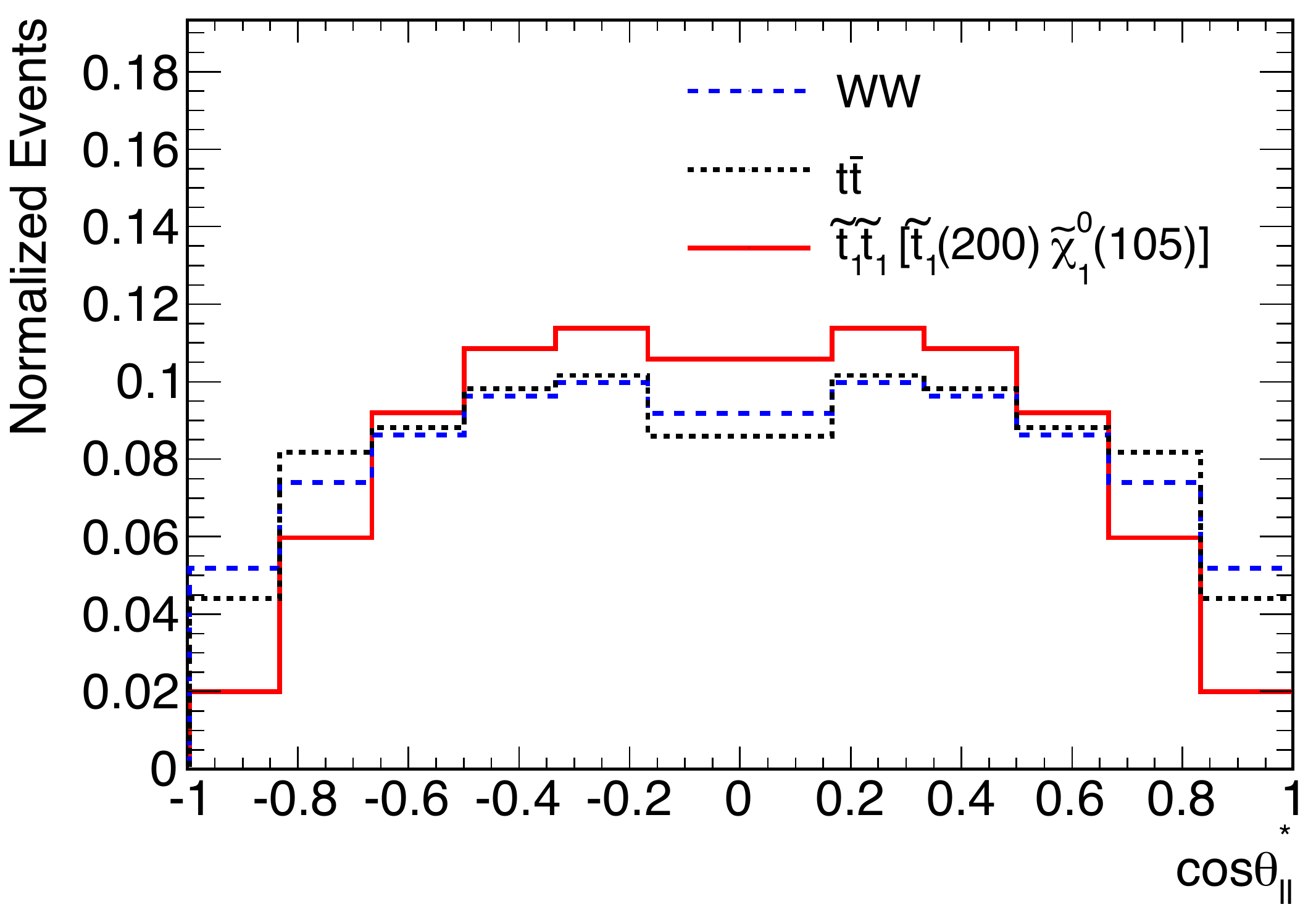}\label{fig:cth0}} 
\caption{ (a) The polar angle, $\cos\theta^*$, of the initially produced $W^+W^-$, $t \bar t$ and stop pairs in the center-of-mass of the hard process frame. (b) The pseudorapidity difference of the lepton pair, $\cos\theta^*_{\ell\ell}$ eq.~\eqref{eq:costhstar}, for the $W^+W^-$, $t \bar t$ and stop events. \label{fig:costh} }
\end{center}
\end{figure}

To improve discriminating power of $\cos\theta_{\ell\ell}^*$ one should take events with higher center-of-mass energy of the hard process. This can be achieved using a variable defined as~\cite{Konar:2008ei,Robens:2011zm}
\begin{equation}\label{eq:sqrtshat}
\sqrt{\hat{s}}_\text{min} = \sqrt{E^2-P_z^2} + \mathmet \;,
\end{equation}
with $E, P_z$ being the total energy and longitudinal momentum of the reconstructed leptons.\footnote{We use the definition of $\sqrt{\hat{s}}_\text{min}$, where the mass of invisible particles is $m_{\mathrm{inv}}=0$, i.e.\ as one would have in the SM, cf.\ ref.~\cite{Konar:2008ei}.}  The $\sqrt{\hat{s}}_\text{min}$ distributions for $W^+W^-$, $t\bar{t}$ and $\sstop\sstop^*$ are shown in figure~\ref{fig:cosmin}. We find that the cut $\sqrt{\hat{s}}_\text{min} > 150 \gev$ leads to the highest significance for discriminating the $W^+W^-$ and stop signals. Figure~\ref{fig:costh_150} shows $\cos\theta_{\ell\ell}^*$ distributions after this cut.

\begin{figure}[t]
\begin{center}
  \subfigure[]{\includegraphics[clip,width=0.48\textwidth]{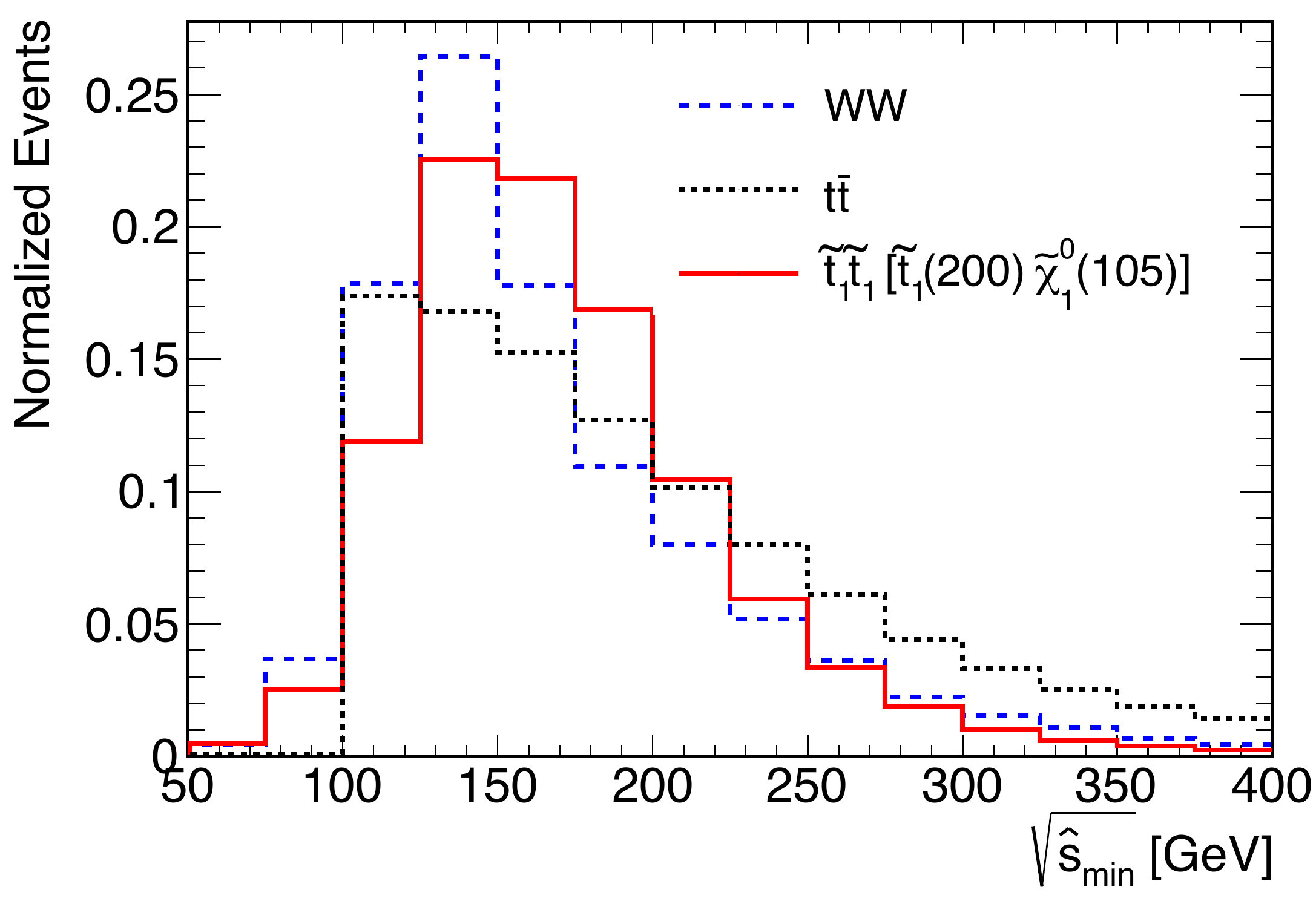}\label{fig:cosmin}} \hspace{0.2cm} 
  \subfigure[]{\includegraphics[clip,width=0.48\textwidth]{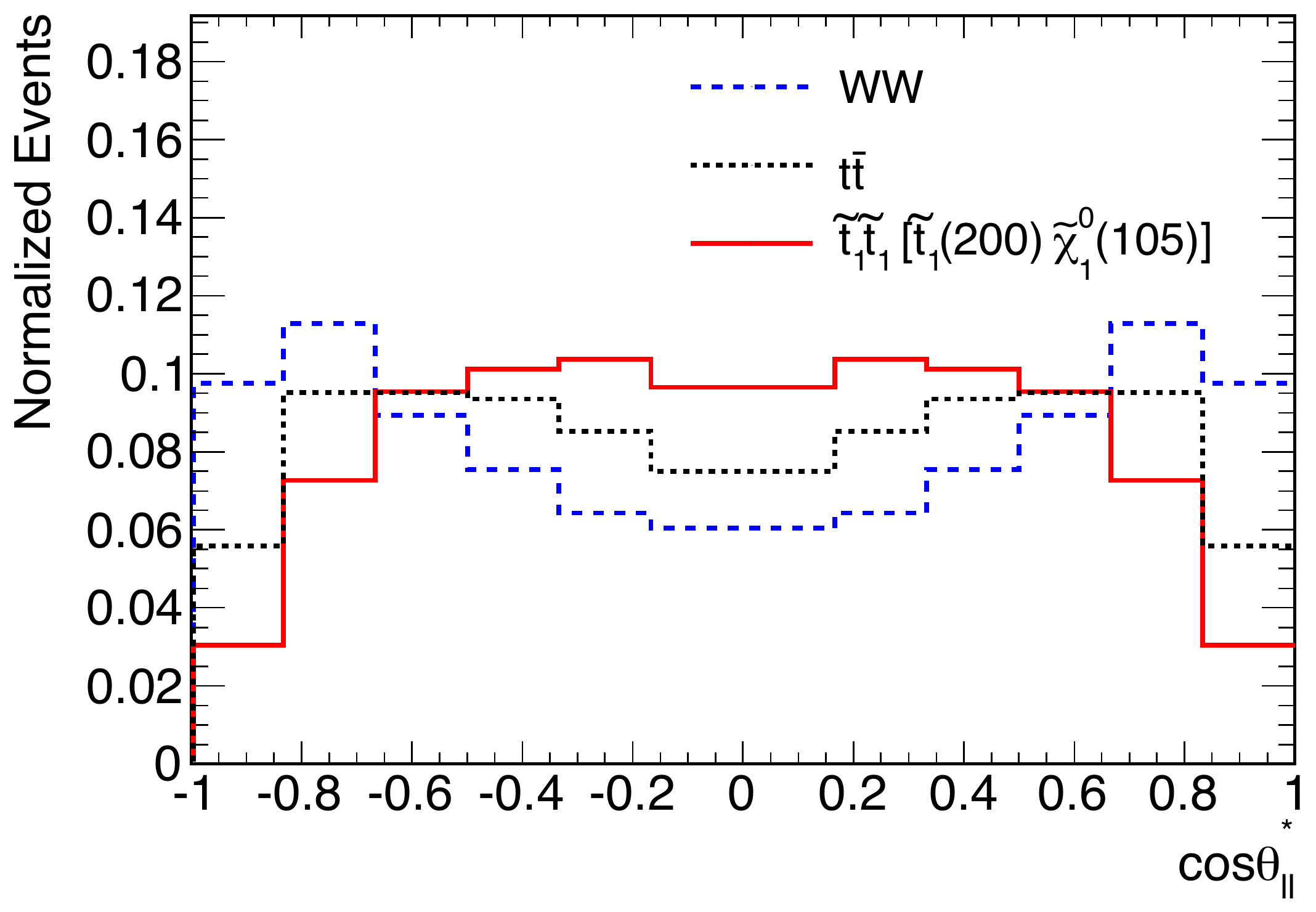}\label{fig:costh_150}} 
\caption{(a) The $\sqrt{\hat{s}}_\text{min}$ distribution for the $W^+W^-$, $t\bar{t}$ and $\sstop\sstop^*$ events. (b) The pseudorapidity difference of the lepton pair, $\cos\theta^*_{\ell\ell}$ eq.~\eqref{eq:costhstar}, for the $W^+W^-$, $t \bar t$ and stops after the selection $\sqrt{\hat{s}}_\text{min} > 150 \gev$.  }
\end{center}
\end{figure}

Finally, we discuss the significance of pinning down the alleged stop signal. We follow here the approach proposed in ref.~\cite{MoortgatPick:2011ix} and define the following asymmetry:
\begin{equation}\label{eq:asymmetry}
\mathcal{A} = \frac{ N(|\cos\theta_{\ell\ell}^*|> 0.5) - N(|\cos\theta_{\ell\ell}^*| < 0.5)}{N_{\mathrm{tot}}}\;,
\end{equation}
where $N(\ldots)$ is the number of events fulfilling the respective condition. After applying the CMS cuts~\cite{Chatrchyan:2013oev} we obtain the values for $\mathcal{A}$ listed in table~\ref{tab:asymmetry}. We compare asymmetry for $W^+W^-$ and stop production at different center-of-mass energies. Clearly, after application of the $\sqrt{\hat{s}}_\text{min} > 150 \gev$ cut we get a better separation of the $W^+W^-$ and SUSY contributions. An additional cut will decrease the number of events, however as can be seen in figure~\ref{fig:cosmin} more so for gauge bosons than for stops. On the other hand, the $t\bar{t}$ contribution is only slightly enhanced. Therefore, we obtain a cleaner sample with a preferable kinematics, so one could expect a better sensitivity.

\begin{table}
\begin{center}
\renewcommand{\arraystretch}{1.}\small
\begin{tabular*}{\textwidth}{@{\extracolsep{\fill} }lrrrr|rrrr} \toprule

 & \multicolumn{4}{c|}{$\mathcal{A}(\sqrt{s} = 8 \tev)$} & \multicolumn{4}{c}{$\mathcal{A}(\sqrt{s} = 14 \tev)$} \\
 $\sqrt{\hat{s}}_\text{min}/$GeV & $WW$ & SM & $\tilde{t}_1 \tilde{t}_1^*$ & SM+$\tilde{t}_1 \tilde{t}_1^*$ & $WW$ & SM & $\tilde{t}_1 \tilde{t}_1^*$ & SM+$\tilde{t}_1 \tilde{t}_1^*$ \\ \midrule 
 $>0$ & $-0.170$ & $-0.157$ & $-0.332$ & $-0.182$ & $-0.163$ & $-0.148$ & $-0.319$ & $-0.219$ \\
 $>150$ & $0.170$ & $0.120$ & $-0.225$ & $0.067$ & $0.197$ & $0.111$ & $-0.210$ & $-0.026$ \\ \bottomrule

\end{tabular*}
\end{center}
\caption{The asymmetry, eq.~\eqref{eq:asymmetry}, for the $W^+W^-$, SM ($W^+W^-$, $t\bar{t}$, $WZ$ and $ZZ$), $\sstop\sstop^*$ and SM with the stop contribution without $\sqrt{\hat{s}}_\text{min}$ cut and after applying the $\sqrt{\hat{s}}_\text{min} > 150 \gev$ requirement. The uncertainty, due to limited MC statistics, is about $0.005$.  \label{tab:asymmetry}}
\end{table}

Figure~\ref{fig:significance} shows the expected significance of measuring a difference in the asymmetry between SM-only (i.e.\ $W^+W^-$ and SM backgrounds: $t\bar{t}$, $WZ$ and $ZZ$) case and SM+$\sstop\sstop^*$, assuming the CMS8 selections. In the asymmetry, some of the important systematic uncertainties (PDFs, scale uncertainties etc.) will cancel out, so for each of the channels it can be reliably estimated with a high accuracy. On the other hand, fraction of events from each of the channels will be prone to systematic uncertainty, that can be calculated using the data of ref.~\cite{Chatrchyan:2013oev}. The systematic uncertainty could be further reduced with more data analysed. The total asymmetry can be now written as
\begin{equation}
 \mathcal{A} = f^{W} \mathcal{A}^{W} + f^{t} \mathcal{A}^{t} + f^{WZ} \mathcal{A}^{WZ} + f^{\tilde{t}} \mathcal{A}^{\tilde{t}} + \ldots\;,
\end{equation}
where $f^i$ and $\mathcal{A}^{i}$ are the fraction of events and the specific asymmetry for each of the signal or background production process and the dots stand for additional background contributions, not included in the present analysis. With this information, one can estimate the systematic uncertainty on the asymmetry to be $\sim 0.01$. Furthermore, we also include statistical uncertainty based on the binomial distribution,
\begin{equation}
\delta(\mathcal{A})_{\mathrm{stat}} = \sqrt{\frac{1-\mathcal{A}^2}{N_{\mathrm{tot}}}}\,. 
\end{equation}
For $10^3$ events this corresponds to $\delta(\mathcal{A})_{\mathrm{stat}}=0.032$ (CMS reported 1111 events in~\cite{Chatrchyan:2013oev}) and scales as $1/\sqrt{N}$ with higher statistics. A clear advantage of using $\sqrt{\hat{s}}_\text{min}$ cut is visible. 
By combining data collected by both ATLAS and CMS, a 3-sigma evidence is possible at $\sqrt{s} = 8 \tev$. At $\sqrt{s} = 14 \tev$, on the other hand, the significance builds up much quicker, providing 5-sigma discrimination with a few$\ifb$ of data. This projection is obtained using the same selections as above with the cross sections rescaled accordingly.

\begin{figure}[t]
\begin{center}
  {\includegraphics[width=0.425\textwidth]{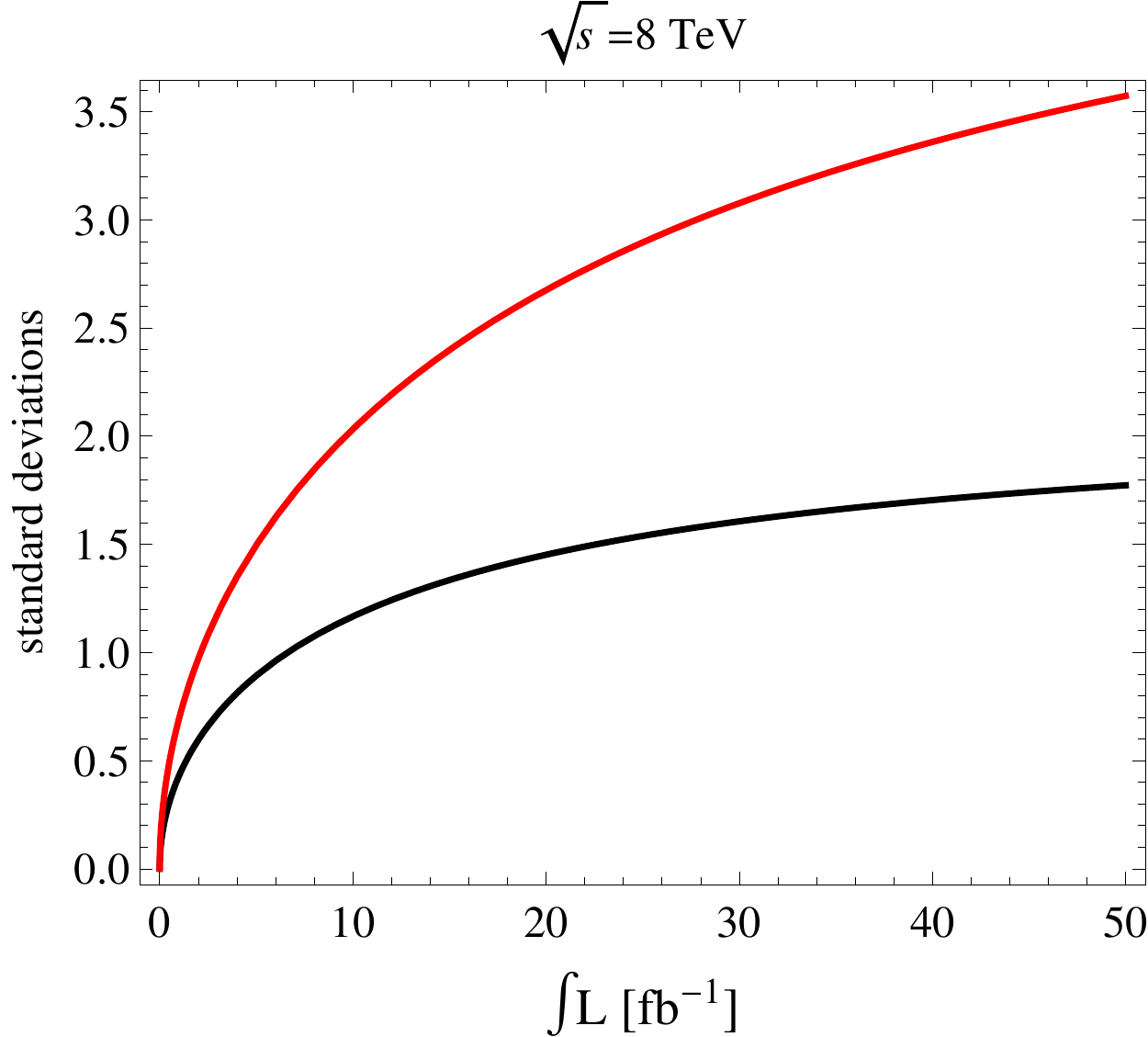}\label{fig:sig8}} \hspace{1cm}
  {\includegraphics[width=0.41\textwidth]{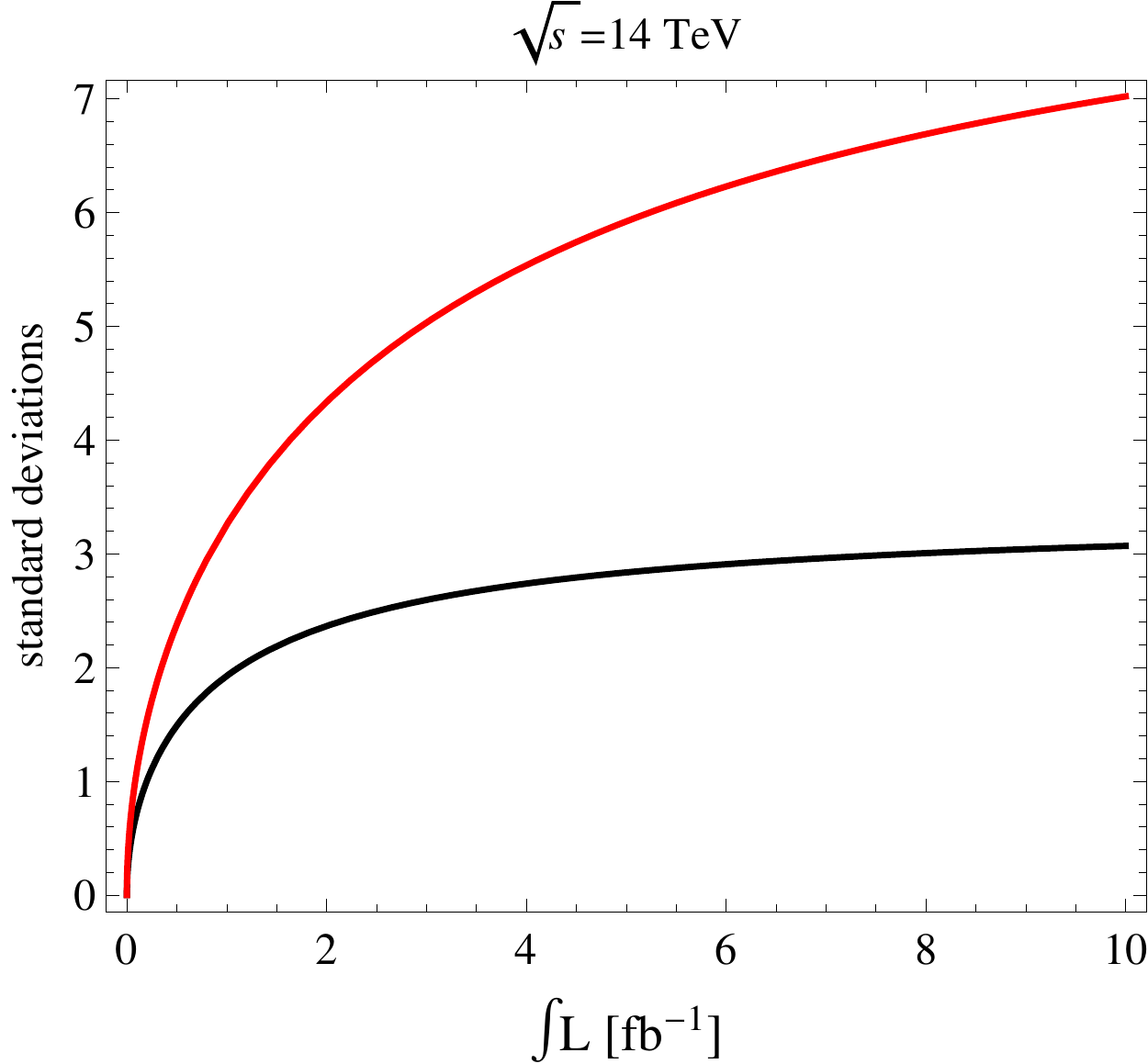}\label{fig:sig14}} 
\caption{The significance of distinguishing the SM-only and SM+$\sstop\sstop^*$ case as a function of an integrated luminosity using the asymmetry eq.~\eqref{eq:asymmetry}. The red curve shows the significance with the cut $\sqrt{\hat{s}}_\text{min} > 150 \gev$, while the black curve without the $\sqrt{\hat{s}}_\text{min}$ cut.  Different $pp$ center-of-mass energies are shown for comparison.  \label{fig:significance} }
\end{center}
\end{figure}

\section{Conclusions\label{conclusions}}
In this paper, we have discussed a possible explanation of the excess in the $W^+W^-$ cross section measurement by the production of supersymmetric partners of top quark. The stop production could provide the right amount of additional signal in the dilepton plus missing transverse energy final state, while the current stop searches could be insensitive.
The large QCD-driven stop cross sections makes it favourable to other possible explanations, like gaugino production within the MSSM. The only requirement in the case of stops is suppression of jet activity, which can be achieved if the mass difference between stop and chargino is small.

We scan the parameter space of the light stop and the lightest neutralino masses to find a region favoured by the present ATLAS and CMS data. The preferred region is localised below $\mstop{1} \sim 250 \gev$ with two branches going down the stop masses. While one of them is excluded by the direct stop searches, the other one remains consistent with the existing limits. It roughly follows a region where $\mstop{1} = \mneu{1} + m_W$. In this region, the kinematic distributions of the stop signal are very similar to the SM $W^+W^-$ pair production distributions. We compare the distributions for a chosen benchmark point, obtaining a good agreement with the results reported by the collaborations. Finally, it can be easily fitted to the Higgs results and the low energy observables.

If the excess is confirmed with a higher significance in a full $8 \tev$ data set, it will be crucial to establish its true nature. Therefore, we have proposed an observable $\cos\theta^*_{\ell\ell}$, based on the angular distributions of the final state leptons, that could help to distinguish between the SM contribution and the genuine stop signal. If the stops are the source of the excess, the full  $8 \tev$ data set could provide a hint of its BSM origin. On the other hand, if the additional data do not confirm the excess our results can be translated to the exclusion limits in yet unconstrained region of the stop parameter space.

A final confirmation of the nature of the excess will require more detailed studies. In particular, one has to show that the new particles decay to the third generation quarks. This task may turn out to be very difficult at the LHC if a mass difference between stop and chargino is very small. In such a case, the final confirmation would require a linear collider with a much higher sensitivity to soft objects. It would be a very interesting scenario for such a machine, with a few new particles in the kinematical reach, allowing for a high precision study of their properties.

\acknowledgments
We thank Masaki Asano, Marco Tonini, Alberto Casas, Jesus Moreno and Bryan Zaldivar for useful discussions.
This work has been partially supported by the MICINN, Spain, under contract FPA2010-17747; Consolider-Ingenio  CPAN CSD2007-00042. 
We thank as well the Comunidad de Madrid through Proyecto HEPHACOS S2009/ESP-1473 and the European Commission under contract PITN-GA-2009-237920.

\bibliographystyle{JHEP}
\bibliography{WW_excess}

\end{document}